# Self-retracting motion of graphite micro-flakes: superlubricity in micrometer scale

Ze Liu [1,2], Jefferson Zhe Liu [3*], Jiarui Yang [2], Yilun Liu [2], Yibing Wang [4], Yanlian. Yang [4], Quanshui Zheng [1,2§]

[1] Institute of Advanced Study, Nanchang University, Nanchang, 330031 China

[2] Department of Engineering Mechanics and Center for Nano and Micro Mechanics, Tsinghua University, Beijing 100084, China

[3] Department of Mechanical and Aerospace Engineering, Monash University, Clayton, VIC 3800, Australia

[4] National Center for Nanoscience and Technology, Beijing 100190, China

**Abstract**

Through experimental study, we reveal superlubricity as the mechanism of self-retracting motion of micrometer sized graphite flakes on graphite platforms by correlating respectively the lock-up or self-retraction states with the commensurate or incommensurate contacts. We show that the scale-dependent loss of self-retractability is caused by generation of contact interfacial defects. A HOPG structure is also proposed to understand our experimental observations, particularly in term of the polycrystal structure. The realisation of the superlubricity in micrometer scale in our experiments will have impact in the design and fabrication of micro/nanoelectromechanical systems based on graphitic materials.

Nano-mechanical devices based on van de Waals forces in multi-walled carbon nanotubes (MWCNT) and HOPG (i.e., multilayered graphenes) have attracted intensive experimental and theoretical studies, owing to their superior properties, e.g., the nearly `freely' motion of inner shell inside the outer shell of a MWCNT [1,2,3], the MWCNT based oscillator with GHz resonance frequency [4], the extremely fast self-retraction motion of graphite flakes in HOPG islands [5] and so on. The role of the interlayer van de Waals interaction in driving the motion of such van de Waals devices has been well recognised and studied by various theoretical analysis and molecular dynamic simulations [3,4,6,7]. On the other hand, the interlayer van de Waals interactions also leads to potential corrugations due to the periodic atomic structures of the graphene layers, and in turn results in the interlayer friction/resistance force. The role of such friction force in the van de Waals micro/nano-mechanical devices, however, is largely overlooked and there is no experimental studies in micrometer scale up to now (except few scanning probe microscope (SPM) experiments with nanoscale sharp tip scanning on top of a graphene [8,9,10,11]). In this Letter, we will reveal the decisive role of such friction force in the van de Waals nano-mechanical devices. Our results show that the superlubricity, as a result of the incommensurate contact of different graphene layers, is the necessary condition for the self-driven motion of CNT/graphene based micro/nano-mechanical devices.

Superlubricity is a phenomenon that friction force vanish or almost vanish when two solid surfaces are sliding over each other [12], and has attracted many attentions [13,14,15,16] since the introduction of the concept [17]. The structural incommensurate between two crystalline solid surfaces leads to the cancelling out of the friction force systematically with the increase of contact area and ultimately leads to the vanishing friction force [13,14]. Laminar materials, such as graphite, MoS2 etc., serve as good systems for superlubricity [15,16], due to their unique structures: atomically smooth contact surface/interface and weak van de Waals interactions between the contact surfaces. Recently, Dienwiebel et. al. occasionally adhered nanometer sized graphite flakes onto a tungsten tip to study the effect of relative rotation of the adhered graphite flakes on the friction behaviour over a graphite platform [12,16]. Their studies showed that the ultra-low friction is indeed caused by the incommensurate contact between the graphite layers. Similar results were observed in the sliding of nano-particles over graphite surface as well [15]. These reported experimental results indicate that the virtually frictionless sliding can occur with the contact area in the nanometer scale. As proposed by Dienwiebel et. al., superlubricity would however break down for sufficiently large contacts as the two contact lattice surfaces are not perfectly rigid [12]. In this Letter, we report the superlubricity in a much larger scale: up to 6 micrometers. This is a step closer to practical applications of superlubricity in MEMS and NEMS.

Let us first take a review of the major observations in the reported self-retracting motion phenomenon of graphite/SiO2 micro-flakes on graphite platforms in our previous work [5]. As sketched in Fig. 1(a), square graphite/SiO2 islands with length 1.0-10.0 μm and height ~200-400nm were fabricated from HOPG (Veeco, ZYH grade) by using the similar technique as proposed in Ref [18] . The island samples were then transferred into a SEM (FEI Quanta 200F) or an optical microscope (OM, HiRox KH-3000) equipped with a micromanipulator MM3A (Kleindiek). Graphite/SiO2 flakes were sheared out from the surface of the islands using the MM3A tip (Fig. 1(b)-(c)), in the same way that layers of graphite rub off from a pencil onto a paper. After the release of MM3A tip, the flakes automatically returned to their original positions on the islands (Fig. 1(c)). The probabilities of self-retraction were reported in Ref. [5] to depend on the length of the graphite islands: 100%, 87%, 33%, and 13%, respectively, for 1.0-2.0, 3.0, 3.5, and 5.0 μm islands of every 15 trials. For the self-retractable flakes, the self-retracting motion can always repeat if no rotation of the flake was forced.

We have shown that the self-retraction motion is mainly driven by the interlayer van de Waals interactions [4,5]. As illustrated in Fig. 1(b), sliding a distance $x$ of the graphite/SiO2 flake will create two exposed graphite basal planes and increase the system free energy $U = 2\gamma L$, where the binding energy $2\gamma = 0.27$ J/m$^2$ is directly determined recently in our experiments [19] and the $L$ denotes the flake length. A consequent restoring force $F_{ret} = - dU/dx = 2\gamma L$ will drive the self-retraction motion. In our previous analysis, the friction force is assumed very small in light of the atomically smooth contact surface and weak van de Waals interactions.

To reveal the role of friction force in the self-retraction motion, we purposely rotate the sheared out graphite/SiO2 top flakes over the bottom flakes with the MM3A tip for our square graphite/SiO2 islands. It is observed that a lock-up state can suddenly appear after a certain rotation angle. Under these lock-up states, not only the self-retractability was lost (Fig. 2(d)), but also shearing and rotating the flake became much harder. If we kept pushing the locked-up islands, the generation of new sliding plane either in the bottom or in the top graphite flake could occur. Repeating this procedure can lead to a series of sheared out flakes as a stack of playing cards (Fig. 1(d)). We find the occurring of the lock-up states were quite regular. Figure 2(a)-(i) summarise the frames of the 'lock-up' states from our *in situ* experiments (supplementary Movie 1), in which the grey frames highlight the bottom graphite flake and the dark-blue arrows align with the same side of the top graphite flake. We translate all the 'lock-up' orientations (i.e., the dark-blue arrows) to a joint point in Fig. 2(j). It is clear to see that the angles between any two neighboured 'lock-up' orientations (shadow areas) are close to 60 degree. The commensurate contacts between graphene planes, i.e., AB stacking, exhibit higher van de Waals friction forces [12,16], which, we believe, could balance the van de Waals restoring force. Thus the good correlation between the observed 'lock-up' states and the commensurate states of the six-fold symmetric graphene (i.e., AB stacking) reveals that the commensurate AB stacking contacts result in the 'lock-up' states of the top flakes. We can conclude that the self-retraction can only occur in the superlubricity state (i.e., incommensurate) in the micrometer scale.

Quantitative measurement on the van de Waals friction forces in the incommensurate/ commensurate states is a challenging task in micrometer scale for experiments. We selected a soft MM3A tip to shear the graphite flake in an optical microscope. We found that in the self-retractable states (i.e., superlubricity), the deformation of the microprobes (Fig. 3(b) and Movie 2) is far smaller than the deformation in the 'lock-up' states (Fig. 3(c) and Movie 3). Roughly, the ratio of the tip deflections in the two cases can be estimated as ~30. This is consistent to experiment reports [12] and theoretical calculations [20]: friction coefficient in incommensurate states is one order magnitude smaller than that in the commensurate states. In addition, we also carry out molecular dynamic calculations with AIREBO force field model [21]. Figure 3(a) shows the friction forces $F_f$ of a rigid square graphene with edge length $L$ over a rigid graphene plane. Different relative rotation angles are tried. The friction force in the incommensurate state (i.e., $\theta = 12°$ and $24°$) is about two orders of magnitude lower than that in commensurate states (A/B stacking). In the A/B stacking, the friction forces is proportional to contact area. We have shown the friction forces with different sheared-out distance $x$ starting from $x = 0$ up to the extreme case $x = 0.98L$. The van de Waals restoring force between the two square graphene layers, $F_{ret} = 2\gamma L$, is also shown in Fig. 3(a). It is clear that incommensurate contact would lead to fully self-retraction of graphite islands of any size, since the friction force is always smaller than the restoring force. Our results also suggest that, in the A/B stacking, the square islands with edge size $L < 10$ nm can also fully self-retract, while islands with $L > 10$ nm should exhibit partial self-retraction because the friction force is smaller than the $F_{ret}$ only for a range of $x$. At typical sheared out distance $x$ in our experiments, i.e., $0.5L$ and $0.667L$, islands bigger than 20nm could not self-retract. Even at the extreme case $x = 0.98L$, the upper limit of the (partially) self-retractable islands is ~450 nm. It confirms our conclusion that self-retraction motion can only occur, in the micrometer scale, under the superlubricity state.

Clearly, the observed self-retraction of our graphite islands should be determined by the structural properties of HOPG. A deep insight of HOPG structure is desired to understand our experiments and to aid the design of novel van de Waals micro/nano-mechanical devices.

Figure 4(a)-(b) show the electron backscatter diffraction (EBSD) of our HOPG sample. Figure 4(a) is the crystal-direction map (IPF mapping image). All the grains have a similar basal plane orientation [0001], which is manifested by the almost same colour in the inverse pole figure (inset). Figure 4(b) depicts the polycrystalline structure, where colours represent different grains. The histogram of grain diameter, which is defined as the mean square root of its area, is shown in the inset of Fig. 4(b). The grain size is distributed within the range of 3-60 μm with the peak about 10 μm. Despite the clear polycrystalline structure in the basal plane demonstrated by our EBSD experiments, the HOPG structure in the depth direction is not clear. Recently Park et.al. estimated the depth of the graphite grains about 5-30nm by using FIB/SEM and high resolution transmission electron microscopy (HRTEM) [22]. Combined with our experimental results, we propose a "stone wall" polycrystalline structure for HOPG and our graphite islands, as shown in Fig. 4(c)-(f), where the dashed lines indicate the grain boundaries. Each brick represents a single grain consisting of multiple graphenes layers in perfect A/B stacking. The contact among different grains is most likely non-A/B stacking, which is indicated in our experiments and will be explained later.

*Apparent controversy of self-retraction and the A/B stacking in HOPG* In most of our self-retraction cases, when we sheared out the top flakes *without rotation*, the flakes can self-retract in all directions (Fig. 1(c)). If we assume perfect A/B stacking in our islands, the self-retraction is impossible because we have proved that A/B stacking will prevent self-retraction in the micrometer scale. Two possible islands structures as shown in Fig. 4(c) and (d) are proposed to resolve the controversy. When cutting a graphite island of edge size 1-2 μm from a HOPG of average grain size ~10 μm, it is possible to have a single grain through the whole islands (Fig. 4(c)). If we assumed that an non-A/B stacking between the grains, the interface of this single grain would be the weakest plane in the islands, where the sliding should occurs. Thus the full self-retraction should take place without rotation. Figure 4(d) represents a less ideal case, in which the smooth and non-A/B stacking interface composes of several grain in the basal plane. We have carried out systematical STM studies on the contact interfaces of graphite islands with self-retractions. One typical result is shown in Fig. 4(g) with height profile slowly varying within 0.5nm over 1 μm. It clearly indicates atomically smoothness of the interface. But we cannot distinguish the two cases in Fig. 4(c) and (d) only from our STM scan. In addition, the multi-layers of grains in HOPG structure as proposed in Fig. 4 implies that the sliding could occur in several difference planes in the islands. Indeed, we observe the formation of a stacks of flakes as shown in Fig. 1(d). The interfaces may not be atomically smooth or perfectly incommensurate, since we observed the self-retracting motion either partially directional, or not fully returned.

*Size dependent loss of the self-retraction* The exposed contact interfaces of several graphite islands without self-retractions are investigated by STM. One typical result is shown in Fig. 4(h), in which the height profile exhibits sharp changes with several nanometers in height. It clearly indicates the appearance of defects like steps or scrolls. The number densities of such surface defects are counted as 0.44 - 2.5 / $\mu m^2$ in five samples. We use two possible structures shown in Fig. 4(e)-(f) to understand the generation of the defects. It is reasonable to expect that graphite islands with edge size close to the grain size in the HOPG should have several grains in the sliding interface. Because of the relative weaker strength at the grain boundary, it is possible that steps could generate (illustrated in Fig. 4(e)), and small grains could break (illustrated in Figure 4(f)) when shearing out the top flakes. These interfacial defects result in significant increase of friction forces and are detrimental to the self-retraction motion. The graphite islands of bigger size should have a higher possibility of such interfacial defects. Indeed, by observing the colour discontinuity of the partially suspended graphite/SiO2 flakes in the optical images, we can roughly estimate the probabilities of the rough surfaces: 0% (out of 24 trials), and 17% (out of 23 trials), 18% (out of 45 trials), 75% (out of 24 trials) and 89% (out of 9 trials) for islands of edge size 1.7, 3.0, 5.0, 7.0 and 10.0 $\mu m$, respectively. This explains the size dependent self-retraction of our islands in experiments.

In summary, we have observed orientation dependence of the self-retracting motion of graphite/SiO2 flake and established a correlation between the self-retraction motion and the incommensurate contacts. We reveal the decisive role of the superlubricity (i.e., incommensurate contacts) in the self-retraction motion of our graphite islands. The realisation of the superlubricity in micrometer scale can have a huge impact on the design and development of the van de Waals nano/micro-mechanical devices. Based on our EBSD and STM measurements and the observed self-retraction motion, we propose a polycrystalline structure of HOPG - "stone wall" structure. We believe the grain boundaries in the HOPG should be the origin for generating the interfacial defects, which leads to the the loss of superlubricity and thus the non-self-retraction in relative larger islands.

**Acknowledgment**

We thank F. Grey, Y. Cheng, Z.P. Xu, Z.H. Li, H.W. Zhu and X. Yang for their helpful discussions. Q.S.Z. acknowledges the financial support from NSFC through Grant No. 10832005, the National Basic Research Program of China (Grant No. 2007CB936803), and the National 863 Project (Grant No. 2008AA03Z302). J. Z. Liu thanks the support of new staff grant 2010 and small grant 2011 from engineering faculty at Monash university.

§ Electronic address: zhengqs@tsinghua.edu.cn

* Electronic address: zhe.liu@monash.edu

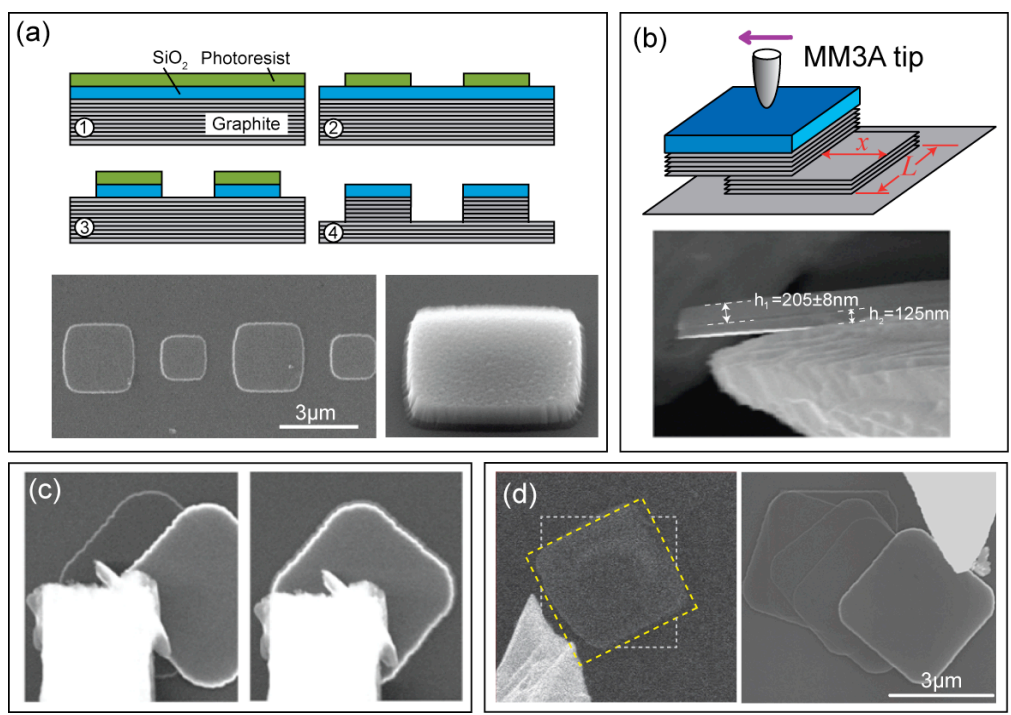

Figure 1. (colour online) (a) Fabrication procedures of HOPG islands, in which a silicon dioxide (SiO2) film is firstly grown on the freshly cleaved HOPG top surface by plasma enhanced chemical vapour deposition (PECVD), then followed by electron beam lithography and reaction ion etching. Top and side views of the obtained graphite islands are shown. (b) Sketch of shearing out of a top graphite/SiO2 flake from the island by MM3A tip. Side view of the sheared out top graphite/SiO2 flake with thickness ~300nm. (c) The self-retraction of sheared out top flake after the release of MM3A tip. (d) Locked-up states with two and multiple stacks of flakes. (Supplementary Movie 1)

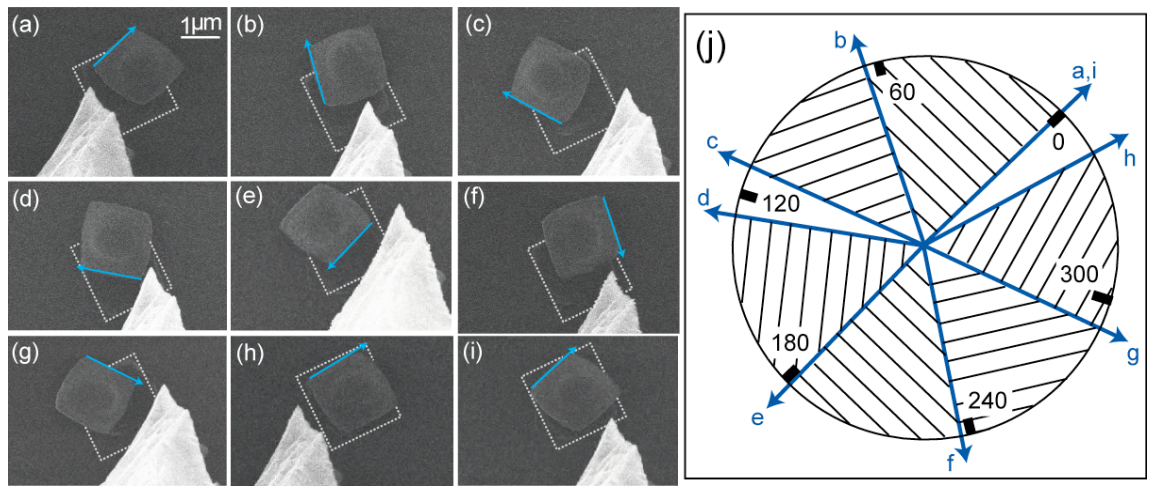

Fig. 2. (colour online) *In situ* manipulation of a graphite/SiO2 flake in a SEM. (a)-(i) show the locked-up states observed in supplementary Movie 1. The dark-blue lines represent the same side of the graphite/SiO2 top flake and the dashed squares (we added for clarity) denote the location of the graphite platform. (j) The joint plot of the 'lock-up' orientations (i.e., the blue lines in (a)-(i)). The angles between two neighboured locked-up orientations are close to 60°.

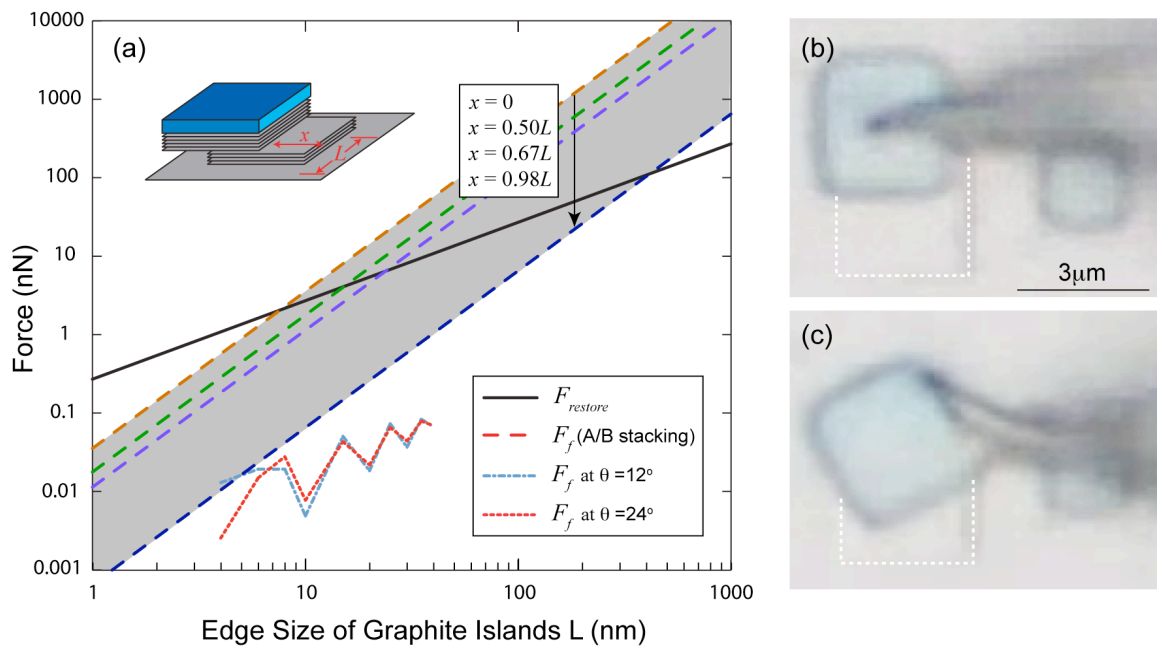

Fig. 3. (colour online) (a) The van de Waals restoring force $F_{res}$ and the friction force $F_f$ in incommensurate (i.e., $\theta = 12°$ and $24°$) and commensurate (i.e., A/B stacking) states as a function of size of a graphite islands (shown in inset). The friction force is a function of sheared out distance $x$, increasing $x$ leading to reduction of contact area and thus lower friction force. Friction forces in commensurate state at four typical $x$ values are shown. (b)-(c) Comparison of deformation of a soft microprobe when moving the top graphite/SiO2 flakes in incommensurate (upper) and commensurate (down) states (Movies 2 and 3).

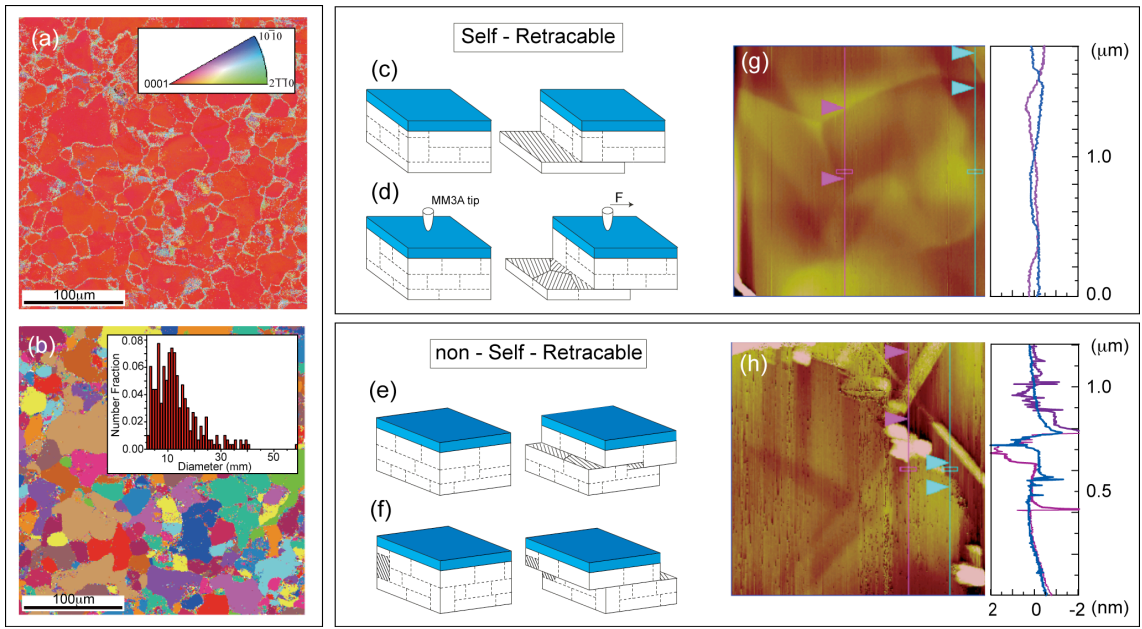

Fig. 4. (color online) (a) Crystal-direction map (IPF mapping) from EBSD with the crystal orientation represented by the inverse pole figure (inset). (b) Auto-grain map from EBSD with colours representing different grains. The histogram of grain diameter, which is defined as the mean square root of its area, is shown in the inset with average size ~10 μm. (c) and (d) represent the two proposed graphite island structures with fully self-retraction motion (see text for details). (e) and (f) depict the two proposed graphite island structures without fully self-retraction motion (see text for details). (g)-(h) STM scan of the typical contact interfaces of the graphite islands with/without fully self-retraction motion, which clearly indicates atomically smooth surface in the self-retraction islands and the appearance of the interfacial defects in the non-self-retractable islands.